\documentclass[12pt]{article}
\usepackage{graphicx}
\begin{document}

\centerline{\Large \bf Penna bit-string model}

\medskip
\centerline{\Large \bf with constant population}
\bigskip

\centerline{S. Moss de Oliveira, P.M.C de Oliveira and J.S. S\'a Martins.}

\bigskip

\noindent Laboratoire PMMH, \'Ecole Sup\'erieure de Physique et de Chimie
Industrielles, 10 rue Vauquelin, F-75231 Paris, Euroland
\medskip

\noindent All visiting from Instituto de F\'{\i}sica, Universidade
Federal Fluminense; Av. Litor\^{a}nea s/n, Boa Viagem,
Niter\'{o}i 24210-340, RJ, Brazil; suzana@if.uff.br, pmco@if.uff.br,  
jssm@if.uff.br.  
\bigskip

Abstract: We removed from the Penna model for biological ageing any 
random killing Verhulst factor. Deaths are due only to genetic diseases 
and the population size is fixed, instead of fluctuating around some 
constant value. We show that these modifications give qualitatively 
the same results obtained in an earlier paper, where the random killings 
(used to avoid an exponential increase of the population) were applied 
only to newborns. 
\medskip



\section{Introduction}

The most successful computational model for age-structured populations is by
far the Penna model \cite{penna}. It's biological support comes from the mutation
accumulation theory of senescence \cite{medawar,gill}. In essence, this theory
relates the evolution of senescence to the action of age-specific deleterious 
mutant alleles and the maintenance of some of these genes by the combined    
effects of mutation and selection pressures \cite{cworth}.

One of the reasons for the Penna model's success relies
on a particularly well-suited computational representation of a genome by
means of a sequence of bits, the bit-string. When grouped into computer words
these strings can be efficiently operated on by very fast logical and bitwise
CPU operations. It has also proved to be flexible enough to
be of value in a number of different problems in population dynamics, running 
from the catastrophic senescence of salmon to the origins of menopause 
(for a review we address the reader to references \cite{book,anais,stauffer}; the 
first one contains also a fortran program to implement the model).  

In order to avoid an exponential increase of the population, the Penna model 
makes use of the so-called Verhulst factor. It is a logistic-type term that 
introduces a mean-field random death probability, independent of the quality 
of the genome.
Its usual expression is $V(Pop) = Pop/Popmax$, where $Pop$ is the total 
population at some time step and $Popmax$ is a parameter of the simulation, 
traditionally called the carrying capacity of the environment. 
Since really random deaths in nature can hardly play any significant
role in population dynamics, this concept has already been criticized
in the literature \cite{cebrat,juja}. Here we modify the model in such a 
way that individuals now die only for genetic reasons. Moreover, the population 
size is fixed, without fluctuations, as the biologists usually prefer.
 
\section{Model with constant population}
 
We start the simulation with $N$ individuals, each one represented 
by a chronological genome. Each of these genomes   
is reduced to the 32 bits of one computer word, where each bit
represents a life-threatening inherited disease. The lifespan is
divided into 32 intervals, each corresponding to one bit position. 
A zero bit means health; a bit set
to one means that starting from that age interval until death, one
additional disease is diminishing the individual's health. $T$ such 
diseases, i.e. $T$ bits set to one in the bitstring from age zero to 
the current age, kill the individual. Those who survive up to age $R$, the 
minimum reproductive age, \underline{may} get offspring, from then on.  
Each offspring differs in $m$ randomly selected bit positions from the parent. 
As usual, we consider only bad mutations which means that if a 
position is selected which has already its bit set to one, then it 
remains one in the offspring genome. 

The reason why we underlined the word ``may'' in the above paragraph is because  
at that point we start to modify the standard model. Now as soon as an individual 
dies, either for reaching the age 32 or the limit $T$ for allowed diseases, a random 
individual is chosen, among all those with age $\ge R$, to generate an offspring. 
In this way we keep the population size constant, without using the Verhulst factor 
already mentioned. This strategy can be interpreted as if, at each iteration, 
we have been chosing a finite sample of an infinite population to observe.     
In the next section we present our results, which are qualitatively the same 
as those obtained earlier by S\'a Martins and Cebrat \cite{juja}. However, 
in their paper they still keep a Verhulst factor acting over the newborns. 
For this reason their population size fluctuates a lot at each iteration. 

\section{Results}

In the figures below we compare the results of the Penna model at constant 
population size with those obtained with the standard model. In order to have  
nearly the same population sizes, we have used for the standard model a carrying 
capacity $Popmax=3,000,000$ individuals, starting with $P_0 = 50,000$ individuals. 
With these parameters the population size of the standard model fluctuates around 
360,000 individuals, after equilibration, while in the constant population model 
it is fixed to 300,000 since the beginning of the simulations.
The rest of the parameters are the same for both: threshold for genetic 
deseases $T=1$, minimum reproduction age $R=8$, mutation rate of newborns $m=1$, 
total number of 1,000,000 timesteps with averages taken over the last 10,000.      
\bigskip

In figure 1 we show the average fraction of indivuals per age. As obtained in 
\cite{juja}, the overall concavity of the constant population curve is different 
from that of the standard one. Also, the maximum life span, the average age 
of an individual and the fraction of population with reproductive live 
(age above 8) are larger for the present model. 
\bigskip 

\begin{figure}[!ht]
\begin{center}
\includegraphics[angle=-90,scale=0.5]{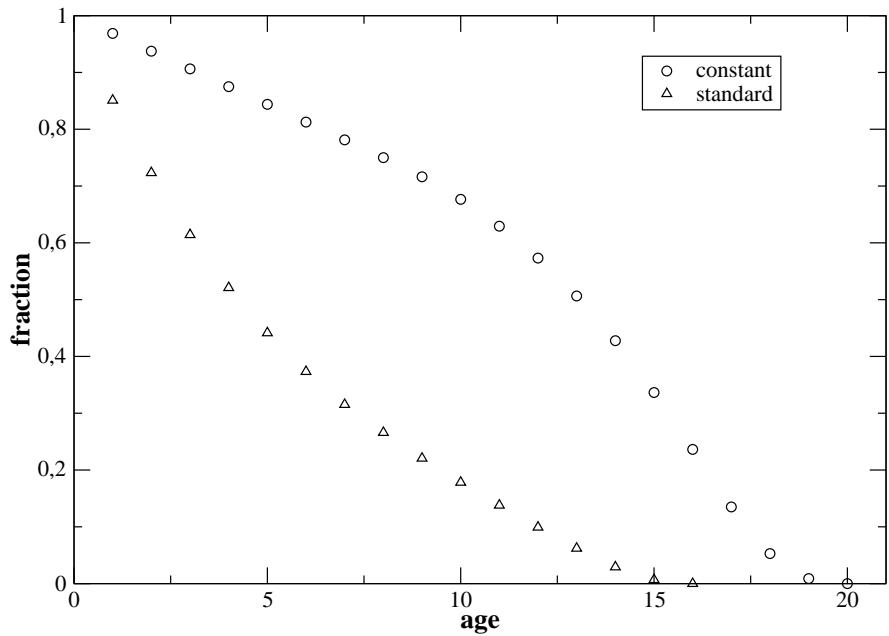}
\end{center}
\caption{Fraction of individuals per age. Circles correspond to the constant population 
model and triangles to the standard one. Parameters are described in the text.
}
\end{figure}

In figure 2 we present the fraction of defected genes (bits 1) for each locus in the 
genome (or age). As expected from the results of the previous figure, the fixation of 
deleterious mutations in the whole population genomes starts earlier with the standard 
model, causing a decrease of the maximum life span, when compared to the constant 
population model.
\bigskip

\begin{figure}[!ht]
\begin{center}
\includegraphics[angle=-90,scale=0.5]{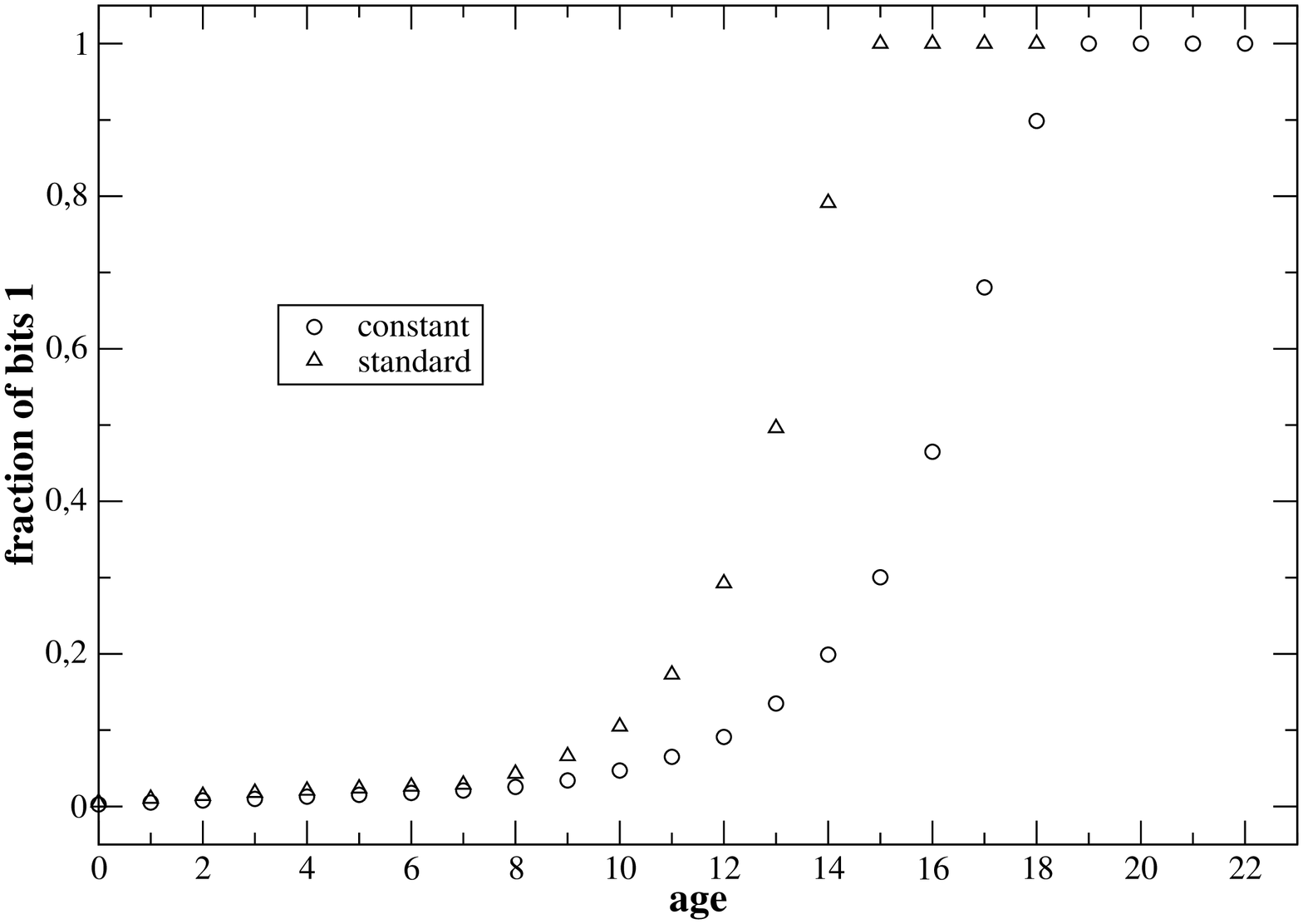}
\end{center}
\caption{Fraction of deleterious alleles (bits set to one) as a function of age.} 
\end{figure}

Figure 3 shows the mortality functions, $q(a)$, defined as: 
$$q(a) = - \ln \left [1 - \frac {D_a} {N_a} \right ] \, \, .$$
$N_a$ is the number of individuals 
with age $a$ and $D_a$ is the number of genetic deaths at age $a$ (which means, 
discounting the deaths provoked by the Verhulst factor in the standard model).
It can be seen that the mortality above the minimum reproduction age 
increases faster in the standard model than in the present one. However, the standard 
model seems to give a better agreement with the Gompertz law \cite{gompertz}, that 
predicts an exponential increase of the mortality above adult ages.

\begin{figure}[!ht]
\begin{center}
\includegraphics[angle=-90,scale=0.5]{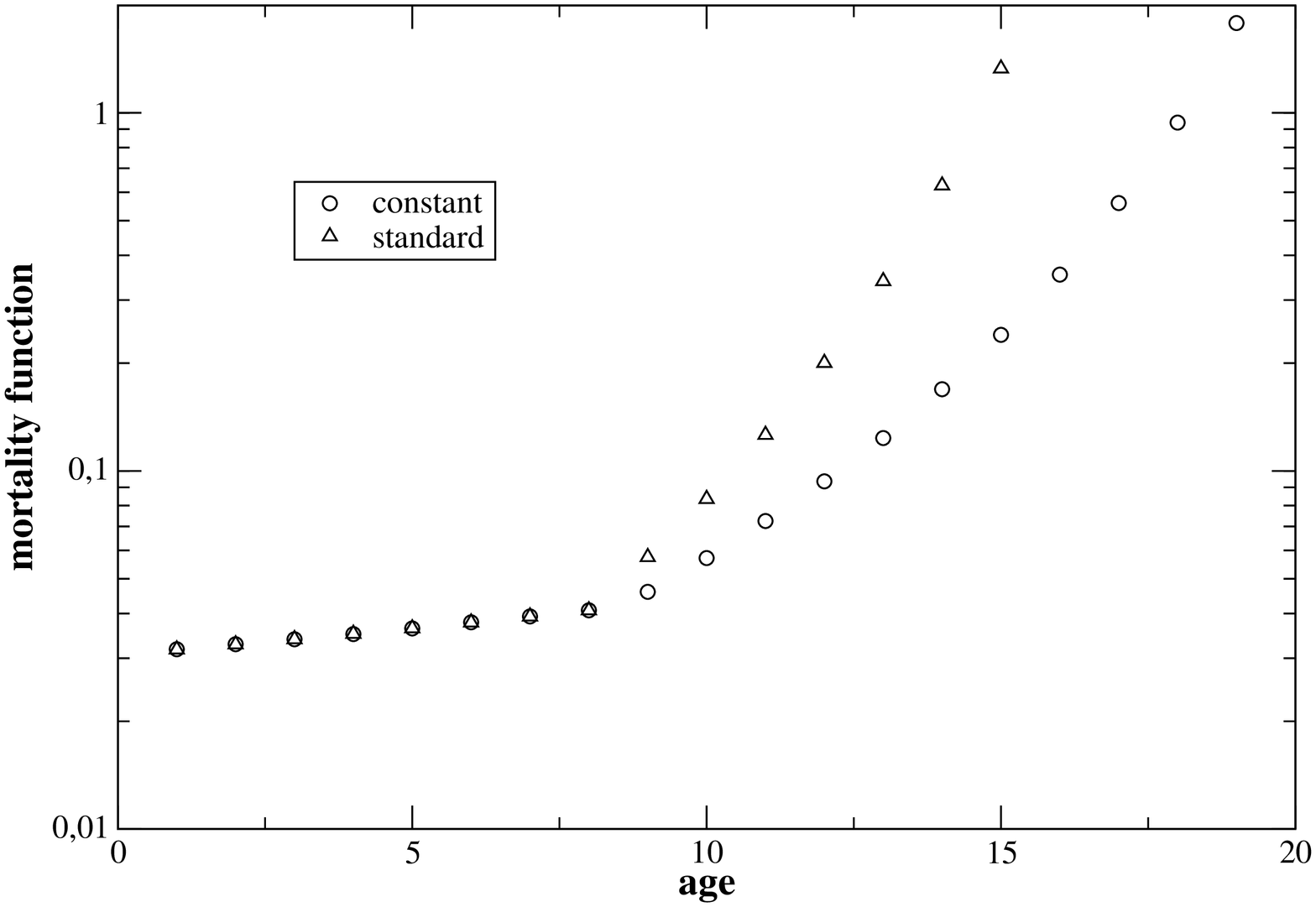}
\end{center}
\caption{Mortality functions in a logarithmic scale.} 
\end{figure}

We have also measured the genetic diversity of the populations, computing the 
difference between each pair of genomes (Hamming distance). The diversity is higher 
for the constant population model, as also obtained in \cite{juja}.  

Real species, of course, can become extinct. Even within a constant population 
approach, by choosing an unrealistic set of parameters (for instance 
a high mutation rate), we observed mutational meltdown through a sudden explosion 
of the birth rate.
 
\section{Conclusions}

We removed from the Penna ageing model any random killing Verhulst factor and 
fixed the population size instead of allowing it to fluctuate around some constant 
value. We observe the same differences between this model and the standard one as 
those pointed out by S\'a Martins and Cebrat: a longer life span, the fixation of 
deleterious alleles starting at older ages and a higher genetic diversity (not shown).  
However, we don't have any fluctuation in the population size at each iteration, 
as they do, once we have a fixed number of individuals since the beginning of the  
simulations. The present model gives a mortality function that increases, for 
ages above the minimum reproduction age, in a slower way than that of the traditional  
Penna model. However, the standard model seems to give a better agreement with the 
Gompertz law of mortality. 

\bigskip

\noindent {\bf Acknowledgements}: To PMMH at ESPCI for the warm hospitality,
to Sorin T\u{a}nase-Nicola for helping us with the computer facilities;
to the Brazilian agencies FAPERJ and CNPq for financial 
support.

\end{document}